\documentclass{ifacconf}

\usepackage{graphicx}      
\usepackage{natbib}        
\usepackage{amsmath,amssymb}
\usepackage{mathtools}
\usepackage{mathrsfs}
\usepackage{graphicx}
\usepackage[colorinlistoftodos]{todonotes}
\usepackage{float}
\usepackage{subfig}
\usepackage{amsfonts}
\usepackage{flafter}
\usepackage{placeins}
\usepackage{bm}
\usepackage{enumerate}
\usepackage{indentfirst}
\usepackage{physics}
\usepackage{ragged2e}
\usetikzlibrary{calc}
\usepackage{import}

\begin{document}

\begin{frontmatter}

\title{Distributed LQR-based observer design for large-scale multi-agent networks}


\author{E.E. Vlahakis~ and~}
\author{G.D. Halikias}

\address{Systems and Control Research Centre, School of Mathematics, Computer Sciences \& Engineering, City, University of London, Northampton Square, London, EC1V 0HB, UK (e-mail: eleftherios.vlahakis@city.ac.uk,~g.halikias@city.ac.uk).}

\begin{abstract}  
In this paper, network of agents with identical dynamics is considered. The agents are assumed to be fed by self and neighboring output measurements, while the states are not available for measuring. Viewing distributed estimation as dual to the distributed LQR problem, a distributed observer is proposed by exploiting two complementary distributed LQR methods. The first consists of a bottom-up approach in which optimal interactions between self-stabilizing agents are defined so as to minimize an upper bound of the global LQR criterion. In the second (top-down) approach, the centralized optimal LQR controller is approximated by a distributed control scheme whose stability is guaranteed by the stability margins of LQR control.  In this paper, distributed observer which minimizes an upper bound of a deterministic performance criterion, is proposed by solving a dual LQR problem using bottom-up approach. The cost function is defined by considering minimum-energy estimation theory where the weighting matrices have deterministic interpretation. The presented results are useful for designing optimal or near-optimal distributed control/estimation schemes.
\end{abstract}


\end{frontmatter}

\section{Introduction}
Multi-agent systems have attracted significant attention of the control community in recent years. Such systems are composed of multiple dynamical systems which are nearly always capable of performing autonomously. The need to solve complex problems and the concept of cooperation require multiple systems working together on common objectives. This is based on interactions between agents which communicate with each other exchanging information regarding their state or local measurements obtained by their sensors. The interactions taking place within the agents' environment constitutes a network of interconnected systems. The number of the interconnections between the agents are limited by the design or practical reasons, resulting in the network topology described by a graph.

Graph theory is widely employed in studying multiple dynamical systems interacting with each other over a communication network. The nodes of the graph correspond to dynamical systems representing agents, while edges represent communication links between them. Problems of regulator, synchronization and consensus are central to the analysis of multi-agent systems and are normally analysed with the tools of spectral graph theory.

Agents are often assumed to have identical dynamics. The coordination of that synchronized behaviour defines the rules and the relationship between agents. These rules/control laws are typically of a distributed nature since shared information is restricted to a neighborhood close to the location of each agent. As a result, even if the whole information is available at the network level, this is not the case for a restricted part of the network containing a single agent, giving rise to the need for distributed control.

The linear quadratic regulator (LQR) is a common method for the control design of systems of this type, due to its well-established properties regarding stability margins. These properties rely strictly on the assumption of the state availability for the purposes of LQR control. From a network perspective, this assumption defines a centralized control scheme, meaning that each agent is able to exchange information with all other agents in the network. In practice, the distributed nature of the network topology can change the properties of the LQR design dramatically. We consider two complementary approaches for solving the distributed LQR problem. The first one is referred to as bottom-up method while the second method as top-down method.

In the bottom-up approach, first, a standard LQR problem is solved at the agent level. This is subsequently complemented by a distributed control scheme which minimizes an upper bound of an aggregate performance criterion involving the whole network. This is achieved by formulating and solving a convex optimization problem involving all agents. This approach is proposed in \cite{postlethwaite}. The main representative of the top-down approach is \cite{borrelli}, where the centralized optimal LQR controller designed for the complete network is approximated by a distributed control scheme by removing a number of control interconnections between agents; the robust stability margins of LQR control guarantee that the resulting system remains stable, provided the number of links removed do not exceed a certain maximum number which depends on the degree of the graph describing agents’ interaction.

The ultimate aim of our work is to generalize the results of LQG control to the distributed case, by assuming that only local output 
measurements are available for feedback. This can be achieved by exploiting the separation principle between estimation (Kalman Filter) and regulation (LQR), suitably generalised to the distributed setting. Note that the issue of robustness is of central importance, as it is well known than in the output-feedback case the guaranteed stability margins of LQR design may be lost. In \cite{ghadami}, a distributed observer design method is proposed by dualising the top-down approach described earlier. Although the work illustrates the structure of the centralized observer and its distributed counterpart arising from the special structure of the weighting matrices chosen in the formulation of the problem, neither the performance criterion nor the weighting matrices are given a clear interpretation, the latter being defined as an issue for future work. In this paper, by using the alternative bottom-up method, we propose suboptimal distributed observer design scheme which minimizes an upper bound of an explicitly defined cost criterion with tuning matrices interpreted deterministically.

\section{NOTATION AND PRELIMINARIES}

We represent the set of real numbers, the set of real-valued vectors of length $m$ and the set of real-valued matrices of $m \times n$ dimensions by $\mathbb{R}$, $\mathbb{R}^{m}$ and $\mathbb{R}^{m \times n}$ respectively.

We consider graphs to represent the interconnections of a network of $N$ agents. A graph $\mathcal{G}$ is defined as $\mathcal{G=(V,E)}$, where $\mathcal{V}$ is the set of nodes (or vertices) $\mathcal{V}=\{1,\cdots, N\}$ and $\mathcal{E}\subseteq \mathcal{V} \times \mathcal{V}$ the set of edges $(i, j)$ with $i\in \mathcal{V}$, $j\in \mathcal{V}$. The degree $d_{j}$ of a graph vertex $j$ is the number of edges which start from $j$. Let $d_{max}(\mathcal{G})$ denote the maximum vertex degree of the graph $\mathcal{G}$. We denote by $\mathbf{A}(\mathcal{G})$ the ($0$, $1$) adjacency matrix of the graph $\mathcal{G}$. Let $\mathbf{A}_{i,j}\in \mathbb{R}$ be its $i$, $j$ element, then $\mathbf{A}_{i,j}=1$ if ($i$, $j$)$\in \mathcal{E}$, $\forall i,j=1,\cdots,N$, $i\neq j$. We will focus on undirected graphs, for which the adjacency matrix $\mathbf{A}(\mathcal{G})$ is symmetric. Let $S(\mathbf{A}(\mathcal{G}))=\{\lambda_{1}(\mathbf{A}(\mathcal{G})),\cdots,\lambda_{N}(\mathbf{A}(\mathcal{G}))\}$ be the spectrum of the adjacency matrix $\mathbf{A}$ associated with an undirected graph $\mathcal{G}$ arranged in nondecreasing semi-order. We define the Laplacian matrix as $L(\mathcal{G})=\mathbf{D}(\mathcal{G})-\mathbf{A}(\mathcal{G})$
where $\mathbf{D}(\mathcal{G})$ is the diagonal matrix of vertex degrees $d_{i}$ (also called the valence matrix). The properties of eigenvalues of Laplacian matrices are strongly related to the structural properties of their associated graphs.

The Kronecker product of $A=[a_{ij}]\in \mathbb{R}^{m \times n}$ and $B=[b_{ij}]\in \mathbb{R}^{p \times q}$ is denoted by $A\otimes B$ and is defined to be the block matrix
$A\otimes B\equiv \begin{pmatrix}
             a_{11}B & \dots  & a_{1n}B \\
             \vdots  & \ddots & \vdots  \\
             a_{m1}B & \dots  & a_{mn}B \end{pmatrix}\in \mathbb{R}^{mp \times nq}$.

Next we outline the setting for the coopertive control: We consider a network of $N$ identical dynamical systems called agents. The dynamical behaviour of every node is described by the differential equation:
\begin{align} \label{eq:post1}
    &\dot{x}_{i}(t)=Ax_{i}(t)+Bu_{i}(t) \nonumber \\
    &x_{i}(0)=x_{i0}
\end{align}
where $x_{i}(t)\in\mathbb{R}^{n}$ and $u_{i}(t)\in\mathbb{R}^{m}$ represent the state and the input vector respectively of the $i^{th}$ individual node. The matrices $A\in\mathbb{R}^{n\times n}$ and $B\in\mathbb{R}^{n\times m}$ are constant and describe the dynamics and the input distribution of the $i^{th}$ individual node/agent respectively. Note that the agents can measure their state and send this information to other agents in the network. This communication scheme is described by a graph $\mathcal{G=(V,E)}$, where $\mathcal{V}$ is the set of the agents ($\mathcal{V}=\{1,\cdots, N\}$) and $\mathcal{E}\subseteq \mathcal{V} \times \mathcal{V}$ the set $(i, j)$ representing the interconnection between agent $i\in \mathcal{V}$ and agent $j\in \mathcal{V}$. We assume that the graph is bidirectional, which means that if the agent $i$ is aware of the state of the agent $j$, then the agent $j$ is aware of the state of the agent $i$. We refer to agent $i$ as $i^{th}$ individual node.

On the grounds that the agents of the network exchange information about their state with each other, we need to model this interaction. We consider the signal $z_{i}(t)$ associated with each node as the sum of the differences between the state of one node and the state of their \textit{neighbours}, taking the form
\begin{equation} \label{eq:post2}
    z_{i}(t)=\sum_{j\in \mathcal{J}_{i}} (x_{i}(t)-x_{j}(t))
\end{equation}
where $i=1,\cdots,N$ and $\mathcal{J}_{i}=\{1,\cdots,N\} \backslash \{i\}$. The subset $\mathcal{J}_{i}$ represents the interconnections of the $i^{th}$ individual node/agent. Intuitively, the signals $z_{i}(t)$ reflects divergence between the state of one agent and the state of agents with which can communicate. Provided state agreement or consensus holds or is established, these signals are equal to zero.

At network level, the system is represented by
\begin{equation} \label{eq:post3}
    \dot{X}(t)=(I_{N}\otimes A)X(t)+(I_{N}\otimes B)U(t)
\end{equation}
where
\begin{equation} \label{eq:post4}
    X(t)=\mathcal{C}ol(x_{1}(t),\cdots,x_{N}(t))
\end{equation}
\begin{equation} \label{eq:post5}
    U(t)=\mathcal{C}ol(u_{1}(t),\cdots,u_{N}(t))
\end{equation}
The signals $z_{i}(t)$ can be represented by
\begin{equation} \label{eq:post6}
    Z(t)=(\mathcal{L}\otimes I_{n})X(t)
\end{equation}
where $\mathcal{L}$ represents the Laplacian of the graph $\mathcal{G}$.

\section{Distributed LQR design}

In this section, we state the regulator problem of a network of $N$ agents where the interactions between them are described by a graph. We consider LQR design where the objective is to define optimal interactions between the agents minimizing an overall cost criterion. Let the overall cost criterion be the cost function at network level
\begin{equation} \label{eq:borre7}
    J(U,X(0))=\int_{0}^{\infty}(X(\tau)^{T}\tilde{Q}X(\tau)+U(\tau)^{T}\tilde{R}U(\tau))d\tau
\end{equation}
where $X(t)$ and $U(t)$ are defined as in (\ref{eq:post4}) and (\ref{eq:post5}) respectively and the matrices $\tilde{Q}$ and $\tilde{R}$ penalize the state and the control at network level.
\subsection{Centralized case}
In this case, we consider that each agent can communicate with every agent in the network and can exchange information about their states. 
Then, the optimal centralized state feedback gain is given by
\begin{equation} \label{eq:borre11}
    \tilde{K}=-\tilde{R}^{-1}\tilde{B}^{T}\tilde{P}
\end{equation}
where the $\tilde{P}$ is the symmetric positive definite solution of the following Algebraic Riccati Equation (ARE):
\begin{equation} \label{eq:borre12}
    \tilde{A}^{T}\tilde{P} + \tilde{P}\tilde{A}-\tilde{P}\tilde{B}\tilde{R}^{-1}\tilde{B}^{T}\tilde{P}+\tilde{Q}=0
\end{equation}
where
\begin{equation*}
    \tilde{A}=(I_{N}\otimes A),\
    \tilde{B}=(I_{N}\otimes B),
\end{equation*}
and $\tilde{Q}\in\mathbb{R}^{Nn\times Nn}$ and $\tilde{R}\in\mathbb{R}^{Nm\times Nm}$.

\subsection{Distributed case}
In this case, each agent in the network communicates with a limited number of agents as described by the adjacency matrix of the graph described
previously. In order to  define the optimal or near-optimal interactions between the agents, we distinguish two complementary methods:
\paragraph*{Top-down method} \hfill \\
\\ Consider first the centralized problem with cost function (\ref{eq:borre7}) and weighting matrices of the form
\begin{align} \label{eq:borre8}
    &\tilde{Q}=\begin{pmatrix}
                     \tilde{Q}_{11} & \tilde{Q}_{12} & \cdots &\tilde{Q}_{1N} \\
                     \vdots         & \ddots         & \vdots &\vdots         \\
                     \tilde{Q}_{N1} & \cdots        &   \cdots&\tilde{Q}_{NN}
                     \end{pmatrix}
    &\tilde{R}=\begin{pmatrix}
                     R & \textbf{0} & \cdots & \textbf{0} \\
                     \vdots         & \ddots         & \vdots &\vdots         \\
                     \textbf{0} & \cdots        &   \cdots& R
                     \end{pmatrix}
\end{align}
with
\begin{align}
    \tilde{Q}_{ii} & =Q_{ii}+\sum_{k=1, \ k\neq i}^{N} \ Q_{ik}, \quad i=1,\cdots,N \nonumber \\
    \tilde{Q}_{ij} & =-Q_{ij}, \quad i,j=1,\cdots,N, \ i\neq j \nonumber \\
    \tilde{R} & = I_{N}\otimes R
\end{align}
and
\begin{subequations}
\begin{eqnarray}
R>0, \ Q_{ii}=Q_{ii}^{T}=Q_{1}\geq 0 \ \forall i \\
Q_{ij}=Q_{ij}^{T}=Q_{ji}=Q_{2}\geq 0 \ \forall i\neq j.
\end{eqnarray}
\end{subequations}
where $Q_{1},Q_{2}\in \mathbb{R}^{n\times n}$ and $R\in\mathbb{R}^{m\times m}$.

The symmetric positive definite solution of (\ref{eq:borre12}) takes the form \cite{borrelli}
\begin{equation}
\tilde{P}=\begin{pmatrix}
            P-(N-1)\tilde{P}_{2} & \tilde{P}_{2} & \cdots & \tilde{P}_{2} \\
            \tilde{P}_{2}       &   P-(N-1)\tilde{P}_{2} & \cdots & \tilde{P}_{2}\\
            \vdots              &   \ddots          & \ddots &      \vdots  \\
            \tilde{P}_{2}       &   \cdots          &   \cdots  &   P-(N-1)\tilde{P}_{2}
        \end{pmatrix}
\end{equation}
where $P$ is the stabilising solution of the ARE at node level
\begin{equation} \label{eq:borre24}
    A^{T}P+PA-PBR^{-1}B^{T}P+Q_{1}=0
\end{equation}
and $\tilde{P}_{2} \in \mathbb{R}^{n \times n}$ is the matrix associated with the ARE
\begin{equation}
\begin{split} \label{eq:borre28}
    (A-XP)^{T} & (-N\tilde{P}_{2})+(-N\tilde{P}_{2})(A-XP)- \\ &-(-N\tilde{P}_{2})BR^{-1}B^{T}(-N\tilde{P}_{2})+NQ_{2}=0
\end{split}
\end{equation}
in which $X=BR^{-1}B^{T}$. Considering the symmetry and the equal weights $Q_1$ on the absolute states of the agents and equal weights $Q_{2}$ on the neighboring state differences, the optimal controller of the centralized scheme is given by
\begin{equation} \label{eq:borre31}
\tilde{K}=\begin{pmatrix}
            K{1} & K_{2} & \cdots & K_{2} \\
            K_{2}       &   K_{1} & \cdots & K_{2}\\
            \vdots              &   \ddots          & \ddots &      \vdots  \\
            K_{2}       &   \cdots          &   \cdots  &   K_{1}
        \end{pmatrix}
\end{equation}
Because of the origin of the $P$ and $P_2$, $K_1$ and $K_2$ are functions of $N$, $A$, $B$, $Q_1$, $Q_2$, and $R$.

Then, assuming that the topology of the network is reflected by the a matrix $M\in \mathbb{R}^{N\times N}$ (see \cite{borrelli}), 
a suboptimal LQR distributed controller can constructed by:
\begin{equation} \label{eq:borre39}
        \hat{K}=-I_{N}\otimes R^{-1}B^{T}P+M\otimes R^{-1}B^{T}\tilde{P}_{2}
\end{equation}
such that $\lambda_{i}>\frac{N_{L}}{2}$ and $N_{L}=d_{max}(\mathcal{G})+1$, where $\lambda_{i}(M)\in S(M)\backslash \{0\}$. For further details
and the main properties of this control scheme see \cite{borrelli}.

\paragraph*{Bottom-up method} \hfill \\
\\ Consider a standard LQR problem involving one self-stabilizing agent. Let $P$ be the symmetric positive definite solution of (\ref{eq:borre24}) and assume that the corresponding optimal controller is given by $K=-R^{-1}B^{T}P$. Then, by considering matrix $\Phi\in\mathbb{R}^{m\times m}$ as design parameter and taking $\mathcal{L}$ to represent the Laplacian matrix of the graph, a state-feedback distributed control scheme
corresponds to a gain matrix of the form:
\begin{equation}\label{eq:pos13}
    \hat{K}=I_{N}\otimes K + \mathcal{L}\otimes \Phi K
\end{equation}
A suboptimal LQR solution, such that an upper bound of an aggregate cost criterion is minimized, can be defined through $\Phi$ by solving a convex optimization problem (see \cite{postlethwaite}).

\section{Distributed Estimation}
Consider network of $N$ agents corresponding to identical dynamical systems in which the (local) state is not available for measurement. Each agent attempts to estimate its state via a local observer (Kalman filter). The agents interact with each other exchanging information about 
the estimation error of their local Kalman filer, according to the interconnection topology described by a graph $\mathcal{G}$ with Laplacian matrix $\mathcal{L}$. The objective is to define an optimal (or near-optimal) distributed observer scheme that estimates the overall state of the network and minimises an aggregate quadratic
cost criterion. Together with the distributed state-feedback LQR gain, the solution to this problem can be used to define distributed dynamic output feedback control
schemes for the network.

In this paper, we use a deterministic Minimum-Energy Estimation (MEE) (see \cite{hespanha}) criterion for the approximate solution of the distributed estimation problem
based on the bottom-up method outlined earlier. The solution defines a suboptimal distributed observer that minimizes an upper bound of the aggregate
performance criterion corresponding to the entire network.

\subsection{Deterministic Minimum-Energy Estimation}
The MEE is the deterministic (but equivalent) version of the standard Kalman filtering estimation problem which relies on stochastic assumptions on the process and
measurement noise signals. We consider a non-exact continuous-time LTI model whose state and output equations are corrupted by a deterministic disturbance and noise signal, respectively:
\begin{equation} \label{eq:hesp4.2}
    \dot{x}=Ax+\bar{B}d, \quad y=Cx+n
\end{equation}
where $d\in\mathbb{R}^{q}$ represents the disturbance and $n\in\mathbb{R}^{m}$ the measurement noise signal, respectively. The objective is to estimate the state trajectory $\bar{x}$ which is consistent with eq. (\ref{eq:hesp4.2}) (i.e. reproduces the exactly the measured signal) and minimises a weighted energy integral defined by the disturbance and noise signals.
Note that since $d$ and $n$ are unknown, for sufficiently large noise/disturbance the equation (\ref{eq:hesp4.2}) does not yield unique solution with respect to $x$.

\paragraph*{Minimum-energy estimation}\hfill \\
\\ The MEE cost function is defined as
\begin{equation} \label{eq:hesp4.4}
    J_{MEE}:= \int_{-\infty}^{t} (n(\tau)^{T}Qn(\tau)+d(\tau)^{T}Rd(\tau)) d\tau
\end{equation}
where $Q\in\mathbb{R}^{m \times m}$ and $R\in \mathbb{R}^{q \times q}$ are symmetric positive definite matrices. Once such a trajectory $\bar{x}$ has been found on the interval $(-\infty,t)$, the MEE is given by $x_{e}(t)=\bar{x}(t)$ (\textit{e} subscript accounts for ``estimation"). The deterministic role of the weighting matrices $Q$ and $R$ tuning the quadratic performance $J_{MEE}$ can be explained as follows \cite{hespanha}:
\begin{itemize}
    \item A large matrix $Q$ means that we impose a large penalty on the noise term, in the sense that the output is corrupted by a small amount of noise, i.e. we have high trust on the measured output. As a result the observer $x_{e}$ should respond fast to changes in $y$.
    \item A large matrix $R$ means that we place a large penalty on the disturbance term, in the sense that the process dynamics are corrupted by a small disturbance signal, i.e. we have high trust on the the past estimates of the trajectory. As a result, the observer should respond slowly to changes in $y$ that are not consistent with the model.
\end{itemize}

\paragraph*{Solution to MEE problem}\hfill \\
\\ A detailed solution to the MEE problem can be found in \cite{hespanha}. Here we present briefly the equations that will be used to obtain the distributed LQR suboptimal observer in the following paragraphs. Considering eq. (\ref{eq:hesp4.2}) we can write the quadratic cost in the form
\begin{align} \label{eq:costMEE}
     J_{MEE}:= \int_{-\infty}^{t} & (C\bar{x}(\tau)-y(\tau))^{T}Q(C\bar{x}(\tau)-y(\tau)) \nonumber \\&+d(\tau)^{T}Rd(\tau)) d\tau
\end{align}

The associated ARE with the cost function $J_{MEE}$ is given by
\begin{equation} \label{eq:hesp4.5}
    (-A^{T})P + P(-A) +C^{T}QC-P\bar{B}R^{-1}\bar{B}^{T}P=0
\end{equation}
where $P$ is symmetric positive definite, resulting in a Hurwitz matrix $-A-\bar{B}R^{-1}\bar{B}^{T}P$. We consider the dual ARE of (\ref{eq:hesp4.5}) as
\begin{equation} \label{eq:hesp4.7}
    AS+ SA^{T} +\bar{B}R^{-1}\bar{B}^{T}-SC^{T}QCS=0
\end{equation}
where $S$ is symmetric positive definite, so that $A-LC$ is Hurwitz in which $L=SC^{T}Q$. Then, the optimal observer with respect to
the cost function (\ref{eq:costMEE}) is given by
\begin{equation} \label{eq:hesp4.6}
    \dot{x}_{e}=(A-LC)x_{e}+Ly
\end{equation}
The condition of existence of the symmetric positive definite solutions $P$ and $S$ of the ARE's (\ref{eq:hesp4.5}) and (\ref{eq:hesp4.7}), respectively, is
\begin{itemize}
    \item The pair $(A,\bar{B})$ is controllable.
    \item The pair $(A,C)$ is observable.
\end{itemize}
From duality, the optimal cost is given by
\begin{equation} \label{eq:costTrace}
    J_{MEE}^*= trace(S)
\end{equation}

%

 \subsection{Distributed Estimation with Bottom-up method and MEE performance}
In this section we use the bottom-up method to minimize an upper bound of the aggregate MEE performance index of the network corresponding to distributed estimation. The network is composed of $N$ identical agents that have access to the estimation error of their local Kalman filter and
also to the estimation error of the Kalman filters of their neighbours. 
Recall that the agent equations are given by:
\begin{equation} \label{eq:nodeEqNoise}
    \dot{x}_{i}=Ax_{i}+\bar{B}d_{i}, \quad y_{i}=Cx_{i}+n_{i}
\end{equation}
where $d_{i}\in\mathbb{R}^{q}$ represents the local disturbance and $n_{i}\in\mathbb{R}^{m}$ the local measurement noise signals, respectively. 
The design procedure goes through several steps involving certain change of coordinates in order to formulate an optimisation problem in a convenient form.

\paragraph*{A. Rotation of coordinates} \hfill \\
\\ We assume without loss of generality that the output matrix $C$ in (\ref{eq:nodeEqNoise}) has the form $C=\begin{pmatrix} 0 & C_{2} \end{pmatrix}$. If this is not the case let $$C=U\begin{pmatrix} \Sigma & 0 \end{pmatrix} V^{T}$$ be the singular value decomposition of the output matrix. Then we change coordinates by the following transformation:
\begin{equation}\label{eq:svdtransf}
    T=\begin{pmatrix} 0 & I_{(n-m)\times (n-m)} \\ I_{m\times m} & 0 \end{pmatrix}V^{T}
\end{equation}  
which puts $C$ into the required form.

\paragraph*{B. Finding optimal observer gain from standard MEE problem} \hfill \\
\\ At this point the gain $L$ is defined at node level (without considering interactions from the network). The associated MEE cost function is given by
\begin{align} \label{eq:costMEEnode}
     J_{MEE}:= \int_{-\infty}^{t} & (C\bar{x}(\tau)-y(\tau))^{T}Q_{1}(C\bar{x}(\tau)-y(\tau)) \nonumber \\&+d(\tau)^{T}Rd(\tau)) d\tau
\end{align}
where $Q_1$ and $R$ are the weighting matrices. The associated dual ARE is
\begin{equation} \label{eq:dualAREabsoluteNode}
    AS+SA^{T} +\bar{B}R^{-1}\bar{B}^{T} -SC^{T}Q_{1}CS=0
\end{equation}
while the optimal gain is:
\begin{equation} \label{eq:gainAbsoluteNode}
    L=SC^{T}Q_{1}
\end{equation}

\paragraph*{C. Change of coordinates (node level)} \hfill \\
\\ This transformation allows the final optimization problem to be solved. First write the Lyapunov matrix $S$ in the form:
\begin{equation} \label{eq:dualpost18}
    S= \begin{pmatrix} S_{11} & S_{12} \\ S_{12}^{T} & S_{22} \end{pmatrix}
\end{equation}
where $S_{11} \in \mathbb{R}^{(n-m)\times (n-m)}$ and $S_{22} \in \mathbb{R}^{(m)\times (m)}$.
The transformation matrix is defined as
\begin{equation} \label{eq:dualpost17}
    \hat{T}:=\begin{pmatrix} I_{m\times m} &  -S_{12}S_{22}^{-1} \\0 & I_ {(n-m)\times (n-m)} \end{pmatrix}
\end{equation}
and the nonsingularity of $S_{22}$ follows from the fact that $S$ is positive-definite. 
The nonsingularity of the transformation matrix $\hat{T}$ is obvious. Under this transformation the set 
$(A, C, S, L, Q_{1})$ becomes $(\hat{A}, \hat{C}, \hat{S}, \hat{L}, \hat{Q}_{1})$ with
\begin{align}
    \hat{A} & =\hat{T}A\hat{T}^{-1} \nonumber \\
    \hat{\bar{B}} & =\hat{T}\bar{B} \nonumber \\
    \hat{C} & =C\hat{T}^{-1} = \begin{pmatrix} 0 & C_{2} \end{pmatrix} \nonumber \\
    \hat{S} & =\hat{T}S\hat{T}^{T} = \begin{pmatrix} S_{11}-S_{12}S_{22}^{-1}S_{12}^{T} & 0 \\ 0 & S_{22} \end{pmatrix} \\
    \hat{L} & = \hat{T}L = \begin{pmatrix} 0 \\ \hat{L}_{2} \end{pmatrix} \\
    \hat{Q}_{1} & =Q_{1} \nonumber \\
    \hat{R} &= R \nonumber
\end{align}

The transformed $\hat{S}$ can easily be obtained by considering the associated dual ARE
\begin{equation} \label{eq:dualAREhat}
    \hat{A}\hat{S}+\hat{S}\hat{A}^{T} +\hat{\bar{B}}R^{-1}\hat{\bar{B}}^{T} -\hat{S}\hat{C}^{T}Q_{1}\hat{C}\hat{S} =0
\end{equation}
Note that matrix $C$ remains invariant ($\hat{C}=C$) under transformation \ref{eq:dualpost17} and from
the block-diagonal structure of $\hat{S}$ it follows that $$\hat{L} = \begin{pmatrix} 0 \\ S_{22}C_{2}^{T}Q_{1} \end{pmatrix} =\begin{pmatrix} 0 \\ \hat{L}_{2} \end{pmatrix}$$ with $L_2\in\mathbb{R}^{m\times m}$ nonsingular.

\paragraph*{D. Design of distributed observer} \hfill \\
\\ The agent/node equation in coordinates \ref{eq:dualpost17} has the form
\begin{align} \label{eq:dualpost21}
    \dot{\hat{x}}_{i}(t) &=\hat{A}\hat{x}_{i}(t)+\hat{\bar{B}}d_{i}(t) \nonumber \\
                y_{i}(t) &=\hat{C}\hat{x}_{i}(t) + n_{i}
\end{align}
with $i=1,\cdots, N$. The distributed observer is defined as
\begin{align} \label{eq:distrObserver}
    \dot{\hat{x}}_{e,i}(t)& =\hat{A}\hat{x}_{e,i}(t) +\hat{L}(y_{i}-\hat{C}\hat{x}_{e,i}(t))+ \nonumber \\
    & +\hat{L}\Phi\sum_{j\in\mathcal{J}_{i}}( (y_{i}-\hat{C}\hat{x}_{e,i}(t)) -(y_{j}-\hat{C}\hat{x}_{e,j}(t)) )
\end{align}
 where $i=1,\cdots,N$ and $\mathcal{J}_{i}=\{1,\cdots,N\} \backslash \{i\}$, $\hat{L}$ is the optimal observer gain associated with cost function (\ref{eq:costMEEnode}), and matrix $\Phi \in \mathbb{R}^{m\times m}$ is the design parameter. From the distributed observer dynamics, it can be seen that the estimate is corrected by the following terms:
\begin{itemize}
    \item local output error $y_{i}-\hat{C}\hat{x}_{e,i}(t)$ of each agent.
    \item relative output error $\sum_{j\in\mathcal{J}_{i}}( (y_{i}-\hat{C}\hat{x}_{e,i}(t)) -(y_{j}-\hat{C}\hat{x}_{e,j}(t)) )$, which accounts for the output error differences between the Kalman filter of the $i$-th node and its neighbouring agents.
\end{itemize}

A distributed observer has optimal-type character only under the condition that a cost function is minimized. The forced structure (\ref{eq:distrObserver}) defines a suboptimal solution which minimizes an upper bound of the networked cost function of the form
\begin{align} \label{eq:detailedMEEhat}
    & J_{MEE}= \int_{\infty}^{t} \bigg( \sum_{i=1}^{N}[(\hat{C}\hat{\bar{x}}_{i}-y_{i})^{T}Q_{ii}(\hat{C}\hat{\bar{x}}_{i}-y_{i})+d_{i}^{T}Rd_{i}] \nonumber \\
    &+\sum_{i=1}^{N}\sum_{j\neq i}^{N}[ ((\hat{C}\hat{\bar{x}}_{i}-y_{i})-(\hat{C}\hat{\bar{x}}_{j}-y_{j}))^{T}Q_{ij}((\hat{C}\hat{\bar{x}}_{i}-y_{i})-\nonumber \\
    & -(\hat{C}\hat{\bar{x}}_{j}-y_{j})) ] \bigg) d\tau
\end{align}
where $Q_{ii}\in \mathbb{R}^{m\times m}$ and $Q_{ij}\in \mathbb{R}^{m\times m}$ are symmetric positive definite weighting matrices that penalize the absolute 
and relative error terms respectively. The MEE cost function at network level is given by
\begin{align} \label{eq:networkMEEhat}
     J_{MEE}= \int_{\infty}^{t} \bigg( &((I_{N}\otimes \hat{C})\hat{\bar{X}}-Y)^{T}((I_{N}\otimes Q_{1})+\nonumber \\
    &+(\mathcal{L}\otimes Q_{2}))((I_{N}\otimes \hat{C})\hat{\bar{X}}-Y) \nonumber \\
    &+D^{T}(I_{N}\otimes R)D\bigg) d\tau
\end{align}
where $\hat{\bar{X}}=(I_{N}\otimes \hat{T})\bar{x}$ represents the aggregate state trajectory of the network in transformed \ref{eq:dualpost17} coordinates that is consistent with the past measurements $Y=\mathcal{C}ol(y_{1},\cdots,y_{N})$ of the network. Here, we have fixed the weighting matrices $Q_{ii}$ and $Q_{ij}$ to $Q_{1}$ and $Q_{2}$ respectively.

The dual ARE associated with the networked performance criterion (\ref{eq:networkMEEhat}) has the form
\begin{align} \label{eq:dualAREnetworkHat}
    (&I_{N} \otimes\hat{A})\hat{S}_{NET}+\hat{S}_{NET}(I_{N}\otimes\hat{A})^{T} + I_{N}\otimes \hat{\bar{B}}R^{-1}\hat{\bar{B}}^{T} -\nonumber \\
    &-\hat{S}_{NET}(I_{N}\otimes \hat{C}^{T}Q_{1}\hat{C}+\mathcal{L}\otimes \hat{C}^{T}Q_{2}\hat{C})\hat{S}_{NET}=0
\end{align}

In transformed \ref{eq:dualpost17} coordinates, the aggregate network equations are given by
\begin{equation} \label{eq:networkEquationhat}
    \dot{\hat{X}}=(I_{N}\otimes \hat{A})\hat{X}+(I_{N}\otimes\hat{\bar{B}})D
\end{equation}
while the aggregate observer equation at network level are given by
\begin{align} \label{eq:networkObserverhat}
    \dot{\hat{X}}_{e}  =((I_{N}\otimes (& A-\hat{L}\hat{C}))-(\mathcal{L}\otimes \hat{L}\Phi\hat{C}))\hat{X}_{e}+ \nonumber \\
    & ((I_{N}\otimes \hat{L})+(\mathcal{L}\otimes \hat{L}\Phi))Y
\end{align}
where $\mathcal{L}$ is the Laplacian of the graph that describes the interconnection topology of the network.

\paragraph*{E. Change of coordinates (network level)} \hfill \\
\\ The transformations defined in this paragraph are performed at network level. The Laplacian matrix $\mathcal{L}$ is symmetric and positive semi-definite. We consider orthogonal matrix $V\in \mathbb{R}^{N \times N}$ formed by the eigenvectors of the Laplacian $\mathcal{L}$. Let $\mathcal{L}=V\Lambda V^{T}$ be the spectral decomposition of the $\mathcal{L}$ and $\Lambda=diag(\lambda_1,\cdots,\lambda_N)$ be the eigenvalue matrix of $\mathcal{L}$.

We consider the following orthogonal state transformation:
\begin{equation} \label{eq:dualpost27}
    \tilde{X}=\tilde{V}_{n}\hat{X}
\end{equation}
where $\tilde{V}_{n}=(V\otimes I_{n})$
thus the transformed terms in the observer equation (\ref{eq:networkObserverhat}) and in the dual ARE (\ref{eq:dualAREnetworkHat}) are given by
\begin{align}
    \tilde{V}_{n}(I_{N}\otimes (A-\hat{L}\hat{C}))\tilde{V}_{n}^{T}&=(I_{N}\otimes (A-\hat{L}\hat{C})) \\
     \tilde{V}_{n} (\mathcal{L}\otimes \hat{L}\Phi \hat{C})\tilde{V}_{n}^{T} &=(\Lambda\otimes \hat{L}\Phi \hat{C})  \\
     \tilde{V}_{n}(I_{N}\otimes (A-\hat{L}))\tilde{V}_{m}^{T}&=(I_{N}\otimes (A-\hat{L})) \\
     \tilde{V}_{n} (\mathcal{L}\otimes \hat{L}\Phi )\tilde{V}_{m}^{T} &=(\Lambda\otimes \hat{L}\Phi \hat{C})  \\
     \tilde{V}_{n}(I_{N}\otimes (\hat{C}^{T}\hat{Q}_{1}\hat{C})\tilde{V}_{n}^{T}&=(I_{N}\otimes \hat{C}^{T}\hat{Q}_{1}\hat{C}) \\
     \tilde{V}_{n} (\mathcal{L}\otimes \hat{C}^{T}\hat{Q}_{2}\hat{C})\tilde{V}_{n}^{T} &=(\Lambda\otimes \hat{C}^{T}\hat{Q}_{2}\hat{C})
\end{align}

Then, the observer equation at network level (\ref{eq:networkObserverhat}) in the new (\textit{tilda}) coordinates takes the form
\begin{align} \label{eq:observerNetworkTilda}
    \dot{\tilde{X}}_{e}  =((I_{N}\otimes (& A-\hat{L}\hat{C}))-(\Lambda\otimes \hat{L}\Phi\hat{C}))\tilde{X}_{e}+ \nonumber \\
    & ((I_{N}\otimes \hat{L})+(\Lambda\otimes \hat{L}\Phi))\tilde{Y}
\end{align}
where $\tilde{Y}=\tilde{V}_{m}Y$. The dual ARE at network level in the \textit{tilda} coordinates takes the form
\begin{align} \label{eq:dualAREnetworkTilda}
    (&I_{N} \otimes\hat{A})\tilde{S}_{NET}+\tilde{S}_{NET}(I_{N}\otimes\hat{A})^{T} + I_{N}\otimes \hat{\bar{B}}R^{-1}\hat{\bar{B}}^{T} -\nonumber \\
    &-\tilde{S}_{NET}(I_{N}\otimes \hat{C}^{T}Q_{1}\hat{C}+\Lambda\otimes \hat{C}^{T}Q_{2}\hat{C})\tilde{S}_{NET}=0
\end{align}

Since matrix $\Lambda$ is diagonal we can decompose the aggregate observer equation (\ref{eq:observerNetworkTilda}) and its dual ARE (\ref{eq:dualAREnetworkTilda}) and represent them at node level as
\begin{align} \label{eq:observerNodeTilda}
    \dot{\tilde{x}}_{e,i}  =(& A-\hat{L}\hat{C}-\lambda_{i}\hat{L}\Phi\hat{C})\tilde{x}_{e,i}+ \nonumber \\
    & ( \hat{L}+\lambda_{i} \hat{L}\Phi))\tilde{y}_{i}
\end{align}
\begin{align} \label{eq:dualAREnodeTilda}
    \hat{A}\tilde{S}_{i}+\tilde{S}_{i}\hat{A}^{T} +& \hat{\bar{B}}R^{-1}\hat{\bar{B}}^{T} -\nonumber \\
    &-\tilde{S}_{i}(\hat{C}^{T}Q_{1}\hat{C}+\lambda_{i}\hat{C}^{T}Q_{2}\hat{C})\tilde{S}_{i}=0
\end{align}
for $i=1,\cdots,N$. For each symmetric positive definite solution $\tilde{S}_{i}$ we associate a MEE cost criterion of the form
\begin{align} \label{eq:nodeMEEhat}
     J_{MEE,i}= \int_{-\infty}^{t}(\hat{C}\tilde{\bar{x}}_{i}-\tilde{y}_{i})&^{T}(Q_{1}+\lambda_{i}Q_{2})(\hat{C}\tilde{\bar{x}}_{i}-\tilde{y}_{i}) \nonumber \\
     &+d_{i}^{T}Rd_{i}) d\tau
\end{align}and the networked MEE cost function
\begin{equation}\label{eq:networkMEEtilda}
    J_{MEE}=\sum_{i=1}^{N} J_{MEE,i}
\end{equation}

Note that the estimate of the observer at the the $i^{th}$ node does not only contain information about the $i^{th}$ agent, but also about the rest of the network.

\paragraph*{F. Optimization problem}\hfill \\
\\ For each decoupled node we consider a symmetric positive definite (s.p.d) Lyapunov matrix $S_{i}$ with the following structure
\begin{equation} \label{eq:dualpost34}
    S_{i}=\begin{pmatrix} S_{i1} & 0 \\0 & S_{2} \end{pmatrix}
\end{equation}
where $S_{i1}\in \mathcal{R}^{(n-m)\times(n-m)}$ and $S_{2}\in \mathcal{R}^{m\times m}$ are s.p.d. The matrix $S_{2}$ is fixed for all $i=1,\cdots,N$. Then, the matrix $\Phi$ is defined by the solution to the following optimization problem.
 \begin{align}
 \Gamma  & = \min \ \sum_{i=1}^{N}trace(S_{i}) \quad s.t. \nonumber \\
    (\hat{A}-\hat{L}\hat{C}-\lambda_{i}&\hat{L}\Phi \hat{C})S_{i}+S_{i}(\hat{A}-\hat{L}\hat{C}-\lambda_{i}L\Phi \hat{C})^{T} + \nonumber \\ \hat{\bar{B}}R^{-1}\hat{\bar{B}}^{T}
    &+(\hat{L}+\lambda_{i}\hat{L}\Phi)Q_{i}^{-1}(\hat{L}+\lambda_{i}L\Phi)^{T} < 0 \label{eq:dualpost35}\\
    \quad & S_{i}>0 \label{eq:dualpost36}
\end{align} for $i=1,\cdots,N$.
Then, $\Gamma$ constitutes an upper bound on the MEE cost (\ref{eq:networkMEEtilda}) at network level and $J_{MEE}\leq\Gamma$.

\paragraph*{G. LMI formulation}\hfill \\
\\ We define an s.p.d matrix:
\begin{equation} \label{eq:dualpost37}
    W_{i}:=S_{i}^{-1}=\begin{pmatrix} W_{i1} & 0 \\ 0 & W_{2} \end{pmatrix}
\end{equation}
where $W_{i1}=S_{i1}^{-1}$ and $W_{2}=S_{2}^{-1}$. We pre and post multiply eq. (\ref{eq:dualpost35}) and we take
\begin{align} \label{eq:dualpost38}
    W_{i}\hat{A}_{c}-\lambda_{i}\hat{Y}\hat{C}&+\hat{A}_{c}^{T}W_{i}-\lambda_{i}\hat{C}^{T}\hat{Y}^{T}+ \nonumber \\
    \begin{pmatrix}  W_{i}\hat{\bar{B}}R^{-1/2} & \vdots & W_{i}\hat{L}+\lambda_{i}\hat{Y}  \end{pmatrix}& \begin{pmatrix} I_{q} & 0 \\ 0 & Q_{i}^{-1}  \end{pmatrix}
    \begin{pmatrix}  R^{-1/2}\hat{\bar{B}}^{T}W_{i} \\ \cdots\\ (W_{i}\hat{L}+\lambda_{i}\hat{Y})^{T} \end{pmatrix} \nonumber \\ \quad<0
\end{align}
where $\hat{Y}=W_{i}\hat{L}\Phi=\begin{pmatrix}  0 \\ W_{2}\hat{L}_{2}\Phi \end{pmatrix}=\begin{pmatrix}  0 \\ \hat{Y}_{2} \end{pmatrix}$ because of the special structure of $\hat{L}$ and $W_{i}$ and $\hat{A}_{c}=\hat{A}-\hat{L}\hat{C}$.
By applying Schur decomposition on (\ref{eq:dualpost38}) and substituting $\Psi=W_{i}\hat{A}_{c}-\lambda_{i}\hat{Y}\hat{C}+\hat{A}_{c}^{T}W_{i}-\lambda_{i}\hat{C}^{T}\hat{Y}^{T}$ the first LMI is defined as
\begin{equation}\label{eq:dualpost39}
    LMI_{1}=\begin{pmatrix}  \Psi & \quad W_{i}\hat{\bar{B}}R^{-1/2} & \quad W_{i}\hat{L}+\lambda_{i}\hat{Y}\\ * & -I_{q} & 0 \\ * & 0 & -Q_{i} \end{pmatrix}
\end{equation}
where $Q_{i}=Q_{1}+\lambda_{i}Q_{2}>0 \quad \forall \lambda_{i}$. We introduce slack-variable matrix $Z_{i}=\begin{pmatrix}  Z_{i1} & 0 \\ 0 & Z_{2} \end{pmatrix}$ compatible with the form of $S_{i}$ and $W_{i}$. Then we force $Z_{i}-W_{i}^{-1}>0$ or $\sum_{i=1}^{N}Trace(S_{i})<\sum_{i=1}^{N}Trace(Z_{i})$. Then, the equivalent minimization problem in terms of LMI formulation is defined as
\begin{equation}\label{eq:convexOptimization}
    \hat{\Gamma} \ = \ min\sum_{i=1}^{N}Trace(Z_{i}) \quad s.t. \begin{cases} LMI_{1}\\LMI_{2}\\W_{i}>0 \end{cases}
\end{equation}where $LMI_{2}=\begin{pmatrix}  -Z_{i} & I_{n} \\ I_{n} & -W_{i} \end{pmatrix}<0$.

This is a convex optimization problem in terms of LMIs variables $W_{i}$, $\hat{Y}$ and $Z_{i}$ and the design parameter $\Phi$ can be obtained from $\Phi=\hat{L}_{2}^{-1}W_{2}^{-1}\hat{Y}_{2}$.

\section{EXAMPLE}
\begin{exmp}
Consider a network of 5 identical \break vehicles/agents moving in a line. The agent equation is given by (\ref{eq:nodeEqNoise}) with $$A=\begin{pmatrix} 0 & 1 \\ -1 & -0.1 \end{pmatrix}, \quad  \bar{B}=\begin{pmatrix} 0 \\ 1 \end{pmatrix},  \quad C=\begin{pmatrix} 1 & 0 \end{pmatrix}$$
We assume cyclic nearest neighbour interconnection between the agents. We construct distributed observer with networked estimate error dynamics given by the matrix \break $A_{e}=I_{N}\otimes (A-LC)-\mathcal{L}\otimes L\Phi C$. Assuming MEE cost criterion and applying bottom-up method we define $L$ and $\Phi$ that render $A_{e}$ Hurwitz. We present the estimate error of the vehicles' position for different performances selecting appropriate weighting matrices $Q_{1}$, $Q_{2}$ and $R$ and arbitrary initial estimate error for each vehicle.
\begin{figure}[h]
\begin{centering}
\includegraphics[scale=0.65]{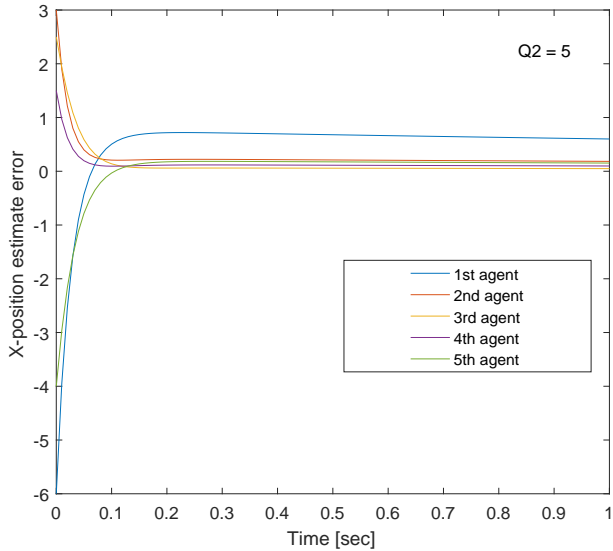}
\end{centering}
\protect\caption{Position estimate error of five identical agents with cyclic nearest neighbour interconnection (low penalty in relative estimation error).} 
\label{fig:1}
\end{figure}

Two cases are provided in this paper. We assume weighting matrices $Q_{1}=10$ and $R=1$ and consider different $Q_{2}$ for each case to illustrate how the relative behavior of the estimates can be shaped by manipulation of the weighting matrices, as explained above. 

The absolute optimal gain matrix $L$ is obtained as \break $L=\begin{pmatrix} 2.6004 \\ 22.8051 \end{pmatrix}$. The optimization problem gives design parameter $\Phi=0.3446$ and performance cost $J=13.4006$ for weighting matrix $Q_{2}=5$ and $\Phi=1.8479$ and $J=13.3401$ for $Q_{2}=25$.
\begin{figure}[h]
\begin{centering}
\includegraphics[scale=0.65]{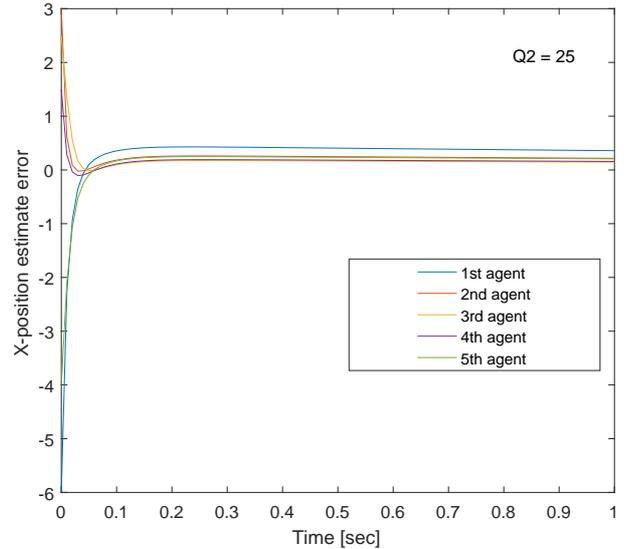}
\end{centering}
\protect\caption{Position estimate error of five identical agents with cyclic nearest neighbour interconnection (high penalty in relative estimation error).} 
\label{fig:2}
\end{figure}

As explained above, the relative sizes of the $Q$, $R$ matrices shift the emphasis between the confidence in the measurements and/or in the model. Moreover, the relative size of $Q_{1}$, $Q_{2}$ can be utilized to shift the emphasis between absolute and relative estimation error respectively. Fig. \ref{fig:1} shows the case with low penalty on relative estimation error while the Fig. \ref{fig:2} indicates faster convergence illustrating the role of the weighting matrix $Q_{2}$. The estimate error system is asymptotically stable, however, it seems to converge slow due to the slowest pole being near the origin. Due to lack of space, more illustrations via simulation results based on this example will be presented at the conference. 

\end{exmp}

\section{Conclusion}
We have introduced a distributed observer of suboptimal character for a network of agents. The procedure requires the solution of a convex optimization problem which depends on the total number of agents. The aggregate cost criterion involves an explicit minimum-energy estimation performance measure and the weighting matrices are given a deterministic interpretation. The overall method depends on the bottom-up approach outlined in the paper and the solution of the distributed LQR problem. By applying the separation principle, the distributed observer obtained using the proposed method, combined with a distributed state-feedback control scheme, can lead to the solution of the dynamic output-feedback distributed control problem. Further work is needed, however, to ensure that the controller obtained in this way has adequate stability margins and can therefore be implemented successfully in practical applications.


\nocite{halikias} 
\nocite{wang}
\nocite{gupta} 
\nocite{cao} 
\nocite{popov}
\nocite{lin} 
\nocite{langbort}
\nocite{bamieh} 
\nocite{massioni}
\nocite{cortes} 
\nocite{tanner} 
\nocite{saber} 
\nocite{beard} 
\nocite{chen} 
\nocite{fax} 

\bibliography{ifacconf}

\end{document}